\documentstyle[twocolumn,aps]{revtex}

\newcommand{\half}{\frac{1}{2}}
\begin{document}
\title{Nonlocal potentials in independent-electron models}
\author{R. K. Nesbet}
\address{
IBM Almaden Research Center,
650 Harry Road,
San Jose, CA 95120-6099, USA}
\date{\today}
\maketitle
\begin{center}For {\em Int.J.Quantum Chem.} \end{center}
\begin{abstract}
This note summarizes the motivation for extending current 
density-functional theory to include nonlocal one-electron potentials, 
and proposes methodology for practical calculations.  The theoretical
model, orbital functional theory, has been shown to be exact in
principle for the general N-electron problem, but must depend in
practice on a parametrized correlation energy functional.  The
discussion here is intended to honor Lee Allen and to bring up to
date some conversations that we began in 1954.
\end{abstract}
\section{Introduction}
The density functional theory (DFT) of Hohenberg, Kohn, and 
Sham\cite{HAK64,KAS65} has become the method of choice for computation
of electronic structure and properties of large molecules and of
solids.  This theory justifies an independent-electron model that is 
considered to be exact in principle for N-electron ground 
states\cite{PAY89,DAG90}.  It is widely assumed that such an exact model
can be expressed in terms of local potential functions for exchange 
and correlation, although evidence to the contrary has existed for some
time\cite{NES01b}.  It has only recently been shown that this locality 
hypothesis is inconsistent with exact theory for more than two electrons
\cite{NES01}. The specific mathematical point is that density functional
derivatives of the Hohenberg-Kohn universal functional cannot be
equivalent to local potential functions, as assumed in established
literature\cite{PAY89,DAG90}, for more than two electrons in a compact
electronic system.  These functional derivatives can be constructed
explicitly\cite{NES01,NES02}, and are found to be G\^ateaux 
derivatives, the generalization to functional analysis of analytic
partial derivatives\cite{BAB92}.  For more than two electrons, they
do not reduce as commonly assumed to simpler Fr\'echet derivatives,
equivalent to multiplicative local potential functions, the 
generalization to functional analysis of analytic total 
derivatives\cite{BAB92}. 
\par This mathematical development explains a number of inconsistencies
or paradoxes that result from assuming the locality 
hypothesis\cite{NES01b}.  Failure of locality for the exchange potential
in a Hartree-Fock variational model of DFT is implied by results of 
optimized effective potential (OEP) calculations going back as far as 
Aashamar, Luke and Talman\cite{ALT78}, confirmed by recent quantitative
tests\cite{CAN01}.  The resulting problems or paradoxes include failure
of DFT linear response theory\cite{PGG96} to agree in the exchange-only
limit with the time-dependent Hartree-Fock theory of Dirac\cite{DIR30}.
This failure can be traced to the locality hypothesis\cite{NES99}.
\par
The most striking paradox affects the kinetic energy.  In contrast to
the locality hypothesis, Kohn and Sham use the kinetic energy operator
of Schr\"odinger for what should be the density functional derivative
of a ground-state kinetic energy functional.  If the locality hypothesis
were a rigorous consequence of variational theory, it should be valid
for the kinetic energy functional, implying for noninteracting electrons
an exact Thomas-Fermi theory (TFT) equivalent to DFT.  However, the
variational equations of these theories are incompatible for more than 
two electrons\cite{NES98}.  The Kohn-Sham equations of DFT are correct 
for atomic shell structure, but the locality hypothesis in TFT is 
inconsistent with the exclusion principle\cite{NES98}. This is evident 
for the lowest $1s2s\,^3S$ state of an atom with two noninteracting 
electrons, perhaps the simplest unequivocal example of this 
inconsistency.  Unless both $1s$ and $2s$ orbital densities are 
independently normalized, the lowest total energy is achieved for two
electrons in the $1s$ state.  Fermi-Dirac statistics require an
independent Lagrange multiplier $\epsilon_i$ for each orbital, while TFT
provides only a single parameter $\mu$ for the total density.
An equivalent inconsistency is implied for interacting 
electrons.  This conclusion has been disputed and defended in recent 
publications\cite{GAL00,HAM01,NES02}.
\par Given a choice between an exact theory and one that can be
expressed entirely in terms of local potential functions, there are
pragmatic arguments supporting either alternative.  The existing
computational methodology of DFT requires use of local 
potential functions.  In the local-density approximation (LDA), which
is entirely justified by variational theory\cite{NES02}, 
first-principles calculations for condensed matter have
had a revolutionary impact on understanding physical properties due
to electronic structure.  However, for magnetically ordered materials,
there is a persistent problem due to the nonphysical electronic
self-interaction, which may require use of nonlocal potentials in
a fully predictive theory.  For molecules, long-range potentials due
to polarization response can become significant in determining 
molecular conformations, but apparently cannot be modeled using only
approximate local correlation potentials.  The optical potentials
required in electron scattering theory are not directly described
by energy-independent local potentials.  
\par With such applications or extensions of existing theory in mind,
the present paper will consider practical issues that arise in
extending existing local-potential methodology to the nonlocal potential
functions appropriate to an ultimately exact independent-electron model.

\section{Variational theory}
The variational theory of independent-electron models is most simply
developed as an orbital functional theory (OFT)\cite{NES01a}. For an
N-electron eigenstate such that $(H-E)\Psi=0$ and any rule $\Psi\to\Phi$
that determines a model or reference state $\Phi$, unsymmetric
normalization $(\Phi|\Psi)=(\Phi|\Phi)=1$ implies that 
$E=(\Phi|H|\Psi)=E_0+E_c$,
where $E_0=(\Phi|H|\Phi)$ is an explicit orbital functional, and 
$E_c=(\Phi|H|\Psi-\Phi)$ defines the correlation energy.
Restricting the discussion to nondegenerate states, $\Phi$ is a Slater 
determinant constructed from occupied orbital functions $\{\phi_i\}$, 
with occupation numbers $n_i=1$.  Spin indices and sums are assumed here
but are suppressed in the notation.  For orthonormal orbital functions, 
$E_0=T+U+V$, where 
\begin{eqnarray}\label{TVops}  
T=\sum_in_i(i|{\hat t}|i);\;V=\sum_in_i(i|v|i),
\end{eqnarray}
for ${\hat t}=-\half\nabla^2$.  Two-electron functionals are defined
for $u=1/r_{12}$ by $U=E_h+E_x$, where
\begin{eqnarray}\label{Uops}
E_h= \half\sum_{i,j}n_in_j(ij|u|ij);\;
E_x=-\half\sum_{i,j}n_in_j(ij|u|ji).
\end{eqnarray}
If ${\cal Q}=I-\Phi\Phi^{\dagger}$,
$E_c=(\Phi|H|\Psi-\Phi)=(\Phi|H|{\cal Q}\Psi)$.
This implies an exact but implicit orbital functional\cite{NES01a} 
\begin{eqnarray}\label{Ecexp}  
E_c=-(\Phi|H[{\cal Q}(H-E_0-E_c-i\eta){\cal Q}]^{-1}H|\Phi),
\end{eqnarray}
for $\eta\to0+$.  
\par Orbital Euler-Lagrange (OEL) equations follow immediately from
standard variational theory\cite{NES02x}, in terms of the 
orbital functional derivatives
\begin{eqnarray}\label{ofds}
\frac{\delta T}{n_i\delta\phi^*_i}={\hat t}\phi_i;\;
\frac{\delta U}{n_i\delta\phi^*_i}={\hat u}\phi_i;\;
\nonumber\\
\frac{\delta V}{n_i\delta\phi^*_i}=v({\bf r})\phi_i;\;
\frac{\delta E_c}{n_i\delta\phi^*_i}={\hat v}_c\phi_i,
\end{eqnarray}
using ${\hat u}=v_h({\bf r})+{\hat v}_x$, where $v_h$ is the classical
Coulomb potential, and ${\hat v}_x$ is the Fock exchange operator.  For
independent subshell normalization, Lagrange terms $[(i|i)-1]\epsilon_i$
are subtracted from the energy functional.  Defining
${\cal G}\phi_i=\frac{\delta E}{n_i\delta\phi^*_i}$,
the variational equation is
\begin{eqnarray}
\int d^3{\bf r}\sum_in_i(\delta\phi^*_i
 \{{\cal G}-\epsilon_i\}\phi_i+cc)=0,
\end{eqnarray}
for unconstrained orbital variations in the usual Hilbert space.  
This implies the OEL equations
\begin{eqnarray}
 \{{\cal G}-\epsilon_i\}\phi_i=0, \;\;i=1,\cdots,N.
\end{eqnarray}
Orbitals of different energy are orthogonal.
Equivalently, the universal functional $F=E-V$ defines
\begin{eqnarray}\label{fdr}
\frac{\delta F}{n_i\delta\phi^*_i}={\cal F}\phi_i=
 \{{\hat t}+{\hat u}+{\hat v}_c\}\phi_i.
\end{eqnarray}
and the OEL equations are
\begin{eqnarray}\label{OEL}
{\cal F}\phi_i=\{\epsilon_i-v\}\phi_i, \;\;i=1,\cdots,N.
\end{eqnarray}
\par Overlap products in $(\Phi|H|\Phi)$ drop out of this argument,
because they either produce a unit factor or are multiplied by a
factor that vanishes.  Explicitly, for noninteracting electrons, the
same OEL equations are obtained if the simplified functional $E_0=T+V$
is replaced by $(\Phi|{\hat T}+{\hat V}|\Phi)$.  
The functional differential
\begin{eqnarray}\label{dprod}
\sum_in_i&&\int d^3{\bf r}[
\delta\phi^*_i({\bf r})\{{\hat t}+v({\bf r})-\epsilon_i\}\phi_i({\bf r})
+cc] \Pi_{j\neq i}(j|j)
\nonumber\\
+\sum_in_i&&
(i|{\hat t}+v({\bf r})-\epsilon_i|i)\delta\Pi_{j\neq i}(j|j).
\end{eqnarray}
must vanish for unconstrained orbital variations about a stationary
state.  It cannot do so unless all Lagrange multipliers $\epsilon_i$ are
included, precluding an equivalent Thomas-Fermi theory.  Variations of
the overlap product are multiplied by a factor that vanishes.

\section{The need for nonlocal potentials}
Hohenberg-Kohn theory can be proven for a
nondegenerate ground state of any version of OFT in which
$E_{xc}=E_x+E_c$ is approximated by a 
parametrized functional of $\rho$\cite{NES00}.  Model
electronic density $\rho=\sum_in_i\rho_i$ is a sum of orbital
subshell densities $\rho_i({\bf r})=\phi^*_i({\bf r})\phi_i({\bf r})$.
Defining ${\hat v}_{xc}\phi_i=\delta E_{xc}/n_i\delta\phi^*_i$,
the corresponding Kohn-Sham theory is determined by the OEL equations
given above.  The still controversial issue of whether or not the 
effective one-electron potentials in these equations can be expressed
as multiplicative local potential functions depends on the relationship
between the orbital functional derivatives of Eqs.(\ref{ofds}) and the
density functional derivatives needed for DFT. 
\par Density functional derivatives must be consistent with the
OEL equations.  A crucial logical point is that because functional
definitions must be extended to include variations that are not
constrained by normalization\cite{NES02x}, variational theory restricted
to normalized ground states\cite{EAE84} cannot determine OEL equations.
A variational derivation of the Kohn-Sham equations must extend the 
universal ground-state functional of Hohenberg and Kohn to density
variations driven by unrestricted variations in the orbital Hilbert 
space.  This extended functional is denoted here by $F_s$.  Orbital
theory simplifies the mathematics, because such variations are necessary
and sufficient to determine the Euler-Lagrange equations\cite{NES02x}.
\par 
The rigorous theory of Englisch and Englisch\cite{EAE84} proves the 
existence of a generic density functional derivative for a nondegenerate
ground state. Although not considered in the original derivation, it is
clear from the OEL Eqs.(\ref{OEL}) that the undetermined constant in
\cite{EAE84}, Eq.(4.1), must have orbital indices in order to be
consistent with the electronic subshell structure implied by Fermi-Dirac
statistics\cite{NES02}.  This is made explicit by Eqs.(\ref{fdr},
\ref{OEL}) here, which imply that a functional derivative exists that
depends on an orbital subshell index.  Evaluated for a ground 
state\cite{NES01,NES02b}, the functional differential
\begin{eqnarray}
\delta F_s &=&\sum_in_i\int d^3{\bf r} 
\{\delta\phi^*_i({\bf r}){\cal F}\phi_i({\bf r})+cc\}
\nonumber\\&=&\sum_in_i\int d^3{\bf r}
\{\epsilon_i-v({\bf r})\}\delta\rho_i({\bf r})
\end{eqnarray}
determines the functional derivative
\begin{eqnarray}
\frac{\delta F_s}{n_i\delta\rho_i}=\epsilon_i-v({\bf r}). 
\end{eqnarray}
Because this depends on an orbital index, it defines a G\^ateaux 
derivative\cite{NES01,BAB92}.  If a Fr\'echet derivative 
were to exist, equivalent to a multiplicative local potential function, 
it would imply the Thomas-Fermi equation,
\begin{eqnarray}
\frac{\delta F_s}{\delta\rho}=\mu-v({\bf r}). 
\end{eqnarray}
Because all $\delta F_s/n_i\delta\rho_i$ must equal 
$\delta F_s/\delta\rho$ if the latter exists, these equations are 
inconsistent unless all $\epsilon_i$ are equal.  Hence a
Fr\'echet derivative cannot exist for more than one electron of each
spin without violating the exclusion principle.  This result is not in
conflict with rigorous analysis\cite{EAE84}, which establishes the
existence of G\^ateaux derivatives in general, but cannot distinguish
between G\^ateaux and Fr\'echet derivatives because normalization is
constrained\cite{NES01}.  For noninteracting electrons, this analysis
applies to the Kohn-Sham kinetic energy functional $T_s$\cite{NES02}.
If independent subshell normalization is enforced, using G\^ateaux
derivatives and the set of Lagrange multipliers $\{\epsilon_i\}$,
a generalized Thomas-Fermi theory can be expressed in terms of partial
densities and is equivalent to the orbital functional theory of
Eqs.(\ref{OEL})\cite{NES02b}.
\par
The mathematical situation is that for more than one electron of either 
spin, the functional $F_s$ cannot be extended to derive Euler-Lagrange 
equations for an unstructured total density $\rho$, because a Fr\'echet 
functional derivative does not exist.  However, because G\^ateaux 
derivatives exist, an extended functional of orbital subshell densities 
does exist.  Independent subshell variations imply Euler-Lagrange 
equations equivalent to the OEL equations, valid for electronic subshell
structure and for the exclusion principle.  For noninteracting 
electrons, $F_s$ reduces to the Kohn-Sham functional $T_s$, with the 
same mathematical properties.  There is no implication that the density 
functional derivative of $E_{xc}$ must be a local potential function
$v_{xc}({\bf r})$, unless this is mandated by the simplified functional
dependence assumed in the local density approximation (LDA).

\section{Nonlocal potentials}
The concept of a G\^ateaux functional derivative establishes a direct 
relationship between orbital-indexed local potential functions and 
nonlocal potentials defined by orbital functional derivatives.  
Using exchange energy as an example, the orbital functional derivative
is $\delta E_x/n_i\delta\phi^*_i={\hat v}_x\phi_i$,
and the functional differential can be expressed as 
\begin{eqnarray} 
\delta E_x&=&
\sum_in_i\int d^3{\bf r}[\delta\phi^*_i{\hat v}_x\phi_i+cc]
\nonumber\\&=&
\sum_in_i\int d^3{\bf r}\frac{\phi^*_i{\hat v}_x\phi_i}{\phi^*_i\phi_i}
  \delta\rho_i.
\end{eqnarray}
This determines a G\^ateaux derivative
\begin{eqnarray}
\frac{\delta E_x}{n_i\delta\rho_i}=
 \frac{\phi^*_i{\hat v}_x\phi_i}{\phi^*_i\phi_i}=v_{xi}({\bf r}),
\end{eqnarray}
valid for arbitrary variations in the orbital Hilbert space.
The technical problem of generalizing Kohn-Sham methodology to an exact
theory reduces to methodology for such indexed local potentials.  The
occurrence of different potentials for different orbital indices can be
compensated during self-consistent iterations by using off-diagonal  
Lagrange multipliers for orthogonalization.  These parameters must
vanish on convergence for closed-shell OEL equations.  In the case
of exchange in atoms, this is a variant of well-established methodology 
for iterative solution of the Hartree-Fock equations\cite{FRO77}. 
Inclusion of an indexed correlation potential, derived from any
parametrized functional $E_c$,
\begin{eqnarray}
v_{ci}({\bf r})=\frac{\phi^*_i{\hat v}_c\phi_i}{\phi^*_i\phi_i}
\end{eqnarray}
is a straightforward generalization.
\par Hartree-Fock methodology based on expansion in atomic Gaussian
basis orbitals is valid in principle, but becomes impractical for
large molecules and solids.  For solids, multiple scattering theory
(MST)\cite{GAB00} has been the method of choice for DFT calculations.
It provides a common methodology for molecules, based on variational
treatment of space-filling atomic cells\cite{NES02c}.  In MST, the 
principal calculation over a range of orbital energies is to solve
the OEL equations within each local atomic cell, or within the
enclosing sphere of each such cell.  Independent representations of a
wave function within each cell are matched across intercell boundaries 
using variational formalism\cite{NES02c}.  Generalization of the
intracell calculations to nonlocal potentials is straightforward,
simply including an indexed parametrized exchange-correlation potential
while integrating Schr\"odinger or semirelativistic equations at
specified energy.   The practical technical problem is thus  
reduced to the treatment of the long-range tails of indexed local
potentials defined within each cell.  In MST, the formalism is greatly
simplified by the fact that atomic basis functions are literally
truncated at cell boundaries\cite{NES02c}.  This means that there
are no intercell overlap contributions.
\par An approximation that may turn out to be both efficient and
accurate is suggested by the MST formalism.  This is a "local nonlocal"
(LNL) model.  True indexed potentials are to be used only within atomic
cells, but are replaced by their asymptotic forms in external cells.
In general, these asymptotic forms arise from electrostatic multipole
potentials, and can be represented as long-range local potentials.
If this is valid, the indexed potentials in each cell are augmented by
an external field potential that is summed over all other cells.
As in the well-known Ewald expansion, this implies substantial
cancellation of fields, including electrostatic screening, and
may justify neglecting such terms during self-consistent iterations,
including them as first-order perturbations after convergence.
\par An important step toward practical methodology for long-range
correlation has been proposed, and tested in an electron scattering
model\cite{NES00a}.  In a general context, the method is to compute,
for each atomic cell, the first-order perturbed orbital functions due
to external multipole fields.  Intercell correlation energy then can
be formulated in terms of these polarization pseudostates, and
response potentials can be constructed that are correct
at long range, but vanish smoothly at the cell origin.  An example
of dipolar response has been carried out in detail.  The implied
polarization potential is used to calculate polarization response
effects in $e-He$ scattering cross sections\cite{NES00a}.

\end{document}